\documentstyle[12pt,a41,epsfig]{article}
\newcommand{\bq}{\begin{equation}}
\newcommand{\eq}{\end{equation}}

\newcommand\ds{\displaystyle}

\newcommand\GeV{\,\mbox{GeV}}
\newcommand\TeV{\,\mbox{TeV}}

\begin{document}
\sloppy
\thispagestyle{empty}
\begin{flushleft}
DESY 97--032\\
{\tt hep-ph/9703287}\\
March 8, 1997\\
Z. Phys. {\bf C} (1997), in print.
\end{flushleft}

\mbox{}
\vspace*{\fill}
\begin{center}
{\LARGE\bf  On the Expectations for Leptoquarks
} \\

\vspace{2mm}
{\LARGE\bf  in the Mass Range of
{\mbox{\boldmath $O(200$}}~GeV)}\\

\vspace{2em}
\large
Johannes Bl\"umlein
\\
\vspace{2em}
{\it DESY -- Zeuthen,}
 \\
{\it Platanenallee 6,
D--15735 Zeuthen, Germany}\\
\end{center}
\vspace*{\fill}
\begin{abstract}
\noindent
A summary is given on the experimental bounds for the couplings and masses
of scalar and vector leptoquarks associated to the first fermion
generation. We investigate to which extent an interpretation of the
recently reported excess of events in the large $x$ and $Q^2$ range at
HERA in terms of single leptoquark production is compatible with other
experimental results.
\end{abstract}
\vspace*{\fill}
\newpage
\noindent
\section{Introduction}
One of the remarkable properties of the
$SU(3)_c \times SU(2)_L \times U(1)_Y$ Standard Model is the cancellation
of the triangle anomalies, which is implied by the relation
\begin{equation}
\label{ein1}
\sum_n Q_{em,n}^2 \left ( Q_L - Q_R \right )_n = 0
\end{equation}
between the electromagnetic and the left and righthanded weak
charges of the leptons and quarks
for each fermion family. This relation is of fundamental importance,
because it renders the Standard Model renormalizable.
Although not being enforced by every possible scenario,
it might be a dynamical consequence of an as yet unknow underlying theory.

In many  extensions of the Standard Model
new bosons, the leptoquarks~$\Phi_{S,V}$,
which carry both lepton and baryon number,
are predicted.
This applies both to Grand Unified Theories~\cite{GUTS} as well as
models based on compositeness~\cite{COMP} or  technicolor~\cite{TC}.
Whereas in Grand Unified Theories leptoquarks emerge as gauge bosons with
masses in the range of $O(M_{\rm GUT} \sim 10^{15} \GeV)$ and couplings
which do not conserve baryon (B) and lepton (L) number, other
approaches contain also leptoquarks
with B- and L-conserving couplings. These particles may
have masses in the range which is accessible at the present
high energy colliders.

Recently,  the H1 experiment~\cite{H1}
           at HERA observed 12 events in the range of large Bjorken $x$
              for $Q^2 > 15000~\GeV^2$  expecting 4.7 events.
The seven excess events of H1 cluster around $\sqrt{x s} \sim 200 \GeV$.
The ZEUS experiment~\cite{ZEUS} found 4 events
                          for $x > 0.55$ and $y > 0.25$, where
     0.9 events are expected.                   Still the
statistics is low and further measurements are needed before firm
conclusions can be drawn. If the observed excess is confirmed
 at
larger statistics, one of the possible explanations could be single
leptoquark production at a mass $M \sim \sqrt{x s}$,~cf.~\cite{H1}.

In this note we summarize the current experimental limits
on the couplings and masses for scalar and vector leptoquarks and
investigate to which extent this interpretation
is compatible with experimental results obtained at other high energy
colliders.

\section{Excluded mass ranges}
Direct searches for leptoquarks  were performed in $e^+e^-$
annihilation-~\cite{EE}, $p\overline{p}$-~\cite{UA,TEV,TEV1},
and $ep$ scattering
experiments~\cite{EPLQ}. The currently most stringent mass limits,
which are independent of the fermionic couplings, were derived by
the TEVATRON experiments (see~\cite{TEV,TEV1,SURV})
searching for leptoquark pair
production~\cite{SCA,VEC1,VEC2,BBK}.
The following  mass ranges are excluded~:
\begin{equation}
\begin{tabular}{lcrrcl}
$M$ &$<$& 143& \GeV&~~~~~&{\rm~1st~generation~scalar~leptoquarks}\\
$M$ &$<$& 141& \GeV&~~~~~&{\rm~2nd~generation~scalar~leptoquarks}\\
$M$ &$<$& 99&  \GeV&~~~~~&{\rm~3rd~generation~scalar~leptoquarks} \\
$M$ &$<$& 170& \GeV&~~~~~&{\rm~3rd~generation~vector~leptoquarks}.
\end{tabular}
\end{equation}
In deriving the above
bounds it was assumed that the branching fractions into the
charged lepton-quark and neutrino-quark final states
are equal as a working hypothesis.
Rigorous theoretical predictions on the branching fractions do not exist.
For leptoquarks with a purely  righthanded fermion coupling, however,
the decay $\Phi \rightarrow q \nu$ is not possible on tree level, leading
to
$Br(\Phi \rightarrow e q) \sim 1$. The
H1 experiment~\cite{H1} observed  a high mass excess for $e + jet$ and
$\nu + jet$ final states. If both effects are interpreted in terms of a
leptoquark signal the branching fraction is
$Br(\Phi \rightarrow e q) < 1$.

A preliminary result was reported by the D0 collaboration from the
analysis of Run Ib~\cite{D0new} yielding
\begin{eqnarray}
M &<& 175 (147)
\GeV~~{\rm~1st~generation~scalar~leptoquarks,}~Br(S
\rightarrow eq) = 1 (0.5)~.
\end{eqnarray}

The above
bound on 3rd generation vector leptoquarks was derived assuming
the minimal vector coupling to the gluon, $\kappa_G =1$, see section~4.
For the case that the coupling of the vector leptoquark pair
to the gluon is of the Yang--Mills type, $\kappa_G = 0$,
the mass range
$M < 225 \GeV$ is excluded.
Note that the bound of $M < 170 \GeV$
becomes weaker considering besides the anomalous coupling
$\kappa_G$ a second anomalous coupling $\lambda_G$, as shown in
ref.~\cite{BBK}.
Mass limits from the TEVATRON data
for vector leptoquarks
associated to the first and second
fermion generation were not presented yet.

\section{Bounds from low energy reactions}
Bounds on the leptoquark-fermion couplings
from low energy reactions such as meson decays, meson-antimeson mixing,
lepton decays and others were studied in refs.~\cite{OLDER}--\cite{LEUR}
in detail.
Here the classification of leptoquarks as introduced in ref.~\cite{BRW}
\footnote{Scalar and vector leptoquarks are denoted by $S_I$ and $V_I$,
respectively, where $I$ is the weak isospin.}
was widely used.\footnote{The couplings of leptoquarks emerging in
supersymmetric models with $R$-parity violation, cf. refs.~\cite{RPAR},
to fermion fields are described by a similar Lagrangian, containing only
a few terms.}

For the leptoquarks associated to the first family the bounds
are~\cite{DBC}
\begin{eqnarray}
\label{excl1}
\frac{\lambda}{e} &<& 0.20
\times \left ( \frac{M}{200~\GeV} \right)  ~~~{\rm
 for~the~vector~states}~~V_1,V_{1/2},\\
\label{excl2}
\frac{\lambda}{e} &<& 0.40
\times \left ( \frac{M}{200~\GeV} \right)  ~~~{\rm
for~the~scalars}~~S_0, \tilde{S}_0, \tilde{S}_{1/2}.
\end{eqnarray}
The upper bound on the coupling for the other leptoquarks lies between
these values. Here
$\lambda$ denotes either the lefthanded ($\lambda_L$)
or the righthanded ($\lambda_R$) fermion coupling,
and   $e$ is the  electric charge.
In some cases even either
$\lambda_L \gg \lambda_R$ or
$\lambda_R \gg \lambda_L$ has to be obeyed for $M \sim 200 \GeV$,
cf.~\cite{OLDER}.
Similar results were obtained in refs.~\cite{LEUR}
with a lowest bound of
\begin{eqnarray}
\frac{\lambda}{e} < 0.17
\times \left ( \frac{M}{200~\GeV} \right)  ~.
\end{eqnarray}
Although these bounds are partly affected by low energy hadronic
matrix elements, the above numbers set a scale.
\section{Single leptoquark production in
$ep$
scattering at HERA}
The cross section for single leptoquark production in $ep$ scattering
$\sigma(e+q(\overline{q}) \rightarrow \Phi_{S,V})$
was calculated in refs.~\cite{EP1,BRW,EP3}.
Because the leptoquark widths
\begin{eqnarray}
\label{lqw}
\Gamma_{S} &=& \frac{\lambda_{S}^2}{16 \pi} M_S\nonumber\\
\Gamma_{V} &=& \frac{\lambda_{V}^2}{24 \pi} M_V
\end{eqnarray}
are small compared to the masses $M_{S,V}$ one may work
in the narrow width approximation.  The integrated
leptoquark production cross section for $e^{\pm} q (\overline{q})$
scattering reads~\cite{BRW}
\begin{equation}
\sigma(e q)_{\Phi} = \frac{\pi^2}{s}~\alpha~\left
(\frac{\lambda}{e} \right )^2
q(x = M^2/s,\langle Q^2 \rangle) J(\Phi) b(\Phi)~.
\end{equation}
Here
$\alpha = 1/128.9$ denotes
the fine structure constant,
$q(M^2/s)$ the quark (antiquark) density, and $b(\Phi) =
Br(\Phi \rightarrow eq)$ is
the branching ratio in the production channel. $J(\Phi)$
accounts for the
leptoquark spin with $J(S) = 1$ and $J(V) = 2$.
For $\sqrt{s} = 300.3 \GeV$
the production cross sections are estimated to be
\begin{equation}
\label{estim}
\sigma(e q)_{S,V} = 338~pb~b(\Phi)
\left (\frac{\lambda}{e} \right )^2
\left \{
\begin{array}{ll}
\ds V: & \ds \times 2 \\
\ds S: & \ds \times 1
\end{array} \right.
\left \{ \begin{array}{ll}
\ds u: & \ds \times 0.56 ... 0.25 \\
\ds \overline{u}: & \ds \times 0.005 ... 0.001 \\
\ds d: & \ds \times 0.15 ... 0.05 \\
\ds \overline{d}: & \ds \times 0.014 ... 0.003 \end{array}
\right.
\end{equation}
using the parton densities~\cite{CTEQ3} for
$\langle Q^2 \rangle = 20000 \GeV^2$ and $x = 0.4 ... 0.5$.
The integrated luminosities of the H1 and ZEUS experiment
are $14~pb^{-1}$
and $20~pb^{-1}$, respectively.
If the 7 excess events of the H1 experiment~\cite{H1}
are interpreted in
terms of a leptoquark signal, the respective fermionic couplings
are found to be
\begin{equation}
{\rm H1~:}~~(\lambda_S/e) \sqrt{b}~:
0.06...0.09 (u)~~0.61...1.36 (\overline{u})~~0.11...0.19
 (d)~~0.36...0.79 (\overline{d})
\end{equation}
for leptoquark states produced in $e^+ u(\overline{u})$ or
$e^+ d(\overline{d})$ fusion, respectively.
Here, eq.~(\ref{estim}) was used for an estimate, accounting for an
acceptance of $\varepsilon = 0.8$~\cite{H1}.
The  values for the fermionic couplings $\lambda_V$
for vector leptoquarks are given by $\lambda_V = \lambda_S/\sqrt{2}$
from the above numbers.
The corresponding numbers obtained
for the 3 excess events of the
ZEUS experiment, at an acceptance of $\varepsilon = 0.8$~\cite{ZEUS},
are
\begin{equation}
\lambda_{\rm ZEUS} = 0.55~~\lambda_{\rm H1}.
\end{equation}
The leptoquark couplings are found to lie in the allowed range
$\lambda_S/e < 0.4$  and
$\lambda_V/e < 0.2$, respectively, for the up and down quarks, while
for the antiquarks they are larger and widely
excluded by the bounds
given in eqs.~(\ref{excl1},\ref{excl2}).

Since
for scalar leptoquarks the $y$-distribution is flat, while for
vector leptoquarks it behaves $\propto~(1-y)^2$, cf. ref.~\cite{BRW},
this distribution can in principle be used to get an information on
the spin of the produced state. Although at the current statistics
one cannot get a decisive answer, it is instructive to look on the
$y$ averages. As an example, one obtains for the 7 events of H1
$\langle y \rangle = 0.59 \pm 0.02$ in the range $0.4 < y < 0.9$.
The average $y$-values
for a scalar or vector leptoquark  in this range are
$\langle y \rangle_S = 0.65$ and
$\langle y \rangle_V = 0.55$, respectively, while for the standard
 deep inelastic
$e^+p$ cross section
$\langle y \rangle_{\rm DIS} = 0.54$ is obtained for
comparison. This shows, how difficult a distinction between the different
hypotheses might be, since the average value for the case of vector
leptoquarks is nearly the same as for the deep inelastic sample in this
range of $y$. At a  larger statistics a $\chi^2$-analysis may
yield a definite answer.

Aside of the fusion process
$e^{\pm} q(\overline{q}) \rightarrow \Phi_{S,V}$, leptoquarks
should
also be produced via the processes
$e^{\pm} q(\overline{q}) \rightarrow g \Phi_{S,V}$ and
$e^{\pm} g \rightarrow q (\overline{q}) \Phi_{S,V}$. Since the gluon
density is rapidly falling in the range of larger values of $x$,
the latter reaction yields only a small contribution to the
cross section for large
leptoquark masses. Both processes contain a collinear singularity
in the limit $p_{\perp}(\Phi_{S,V}) \rightarrow 0$, which has to be
regulated for the inclusive cross section. For the differential
distributions a lower cut on $p_{\perp}$  is        applied.
A part of these events would
show an apparent two-jet signature, with one jet at a
mass of $M_{S,V}$. Since the leptoquark and the final state parton
are balanced in $p_{\perp}$, the electron is more difficult to isolate
from the hadronic fragments of the leptoquark both kinematically, and
possibly also
due to $\pi^0$ mesons contained in the jet.
For scalar leptoquark production both processes
were studied in ref.~\cite{EP3} referring to the parton parametrization
of ref.~\cite{EHLQ}. For $M_S = 200 \GeV$ and
$p_{\perp}(\Phi_S) > 5 \GeV$ the integrated cross section for
$e u \rightarrow g \Phi_{S}$ amounts
to $200~\times~\lambda_S^2~pb$, i.e. $1.4/Br(\Phi \rightarrow e u)$
events, for ${\cal L} = 14~pb^{-1}$ and $\lambda_S/e  = 0.075$.

\section{Single and Pair Production of Leptoquarks at TEVATRON}
The single--production of leptoquarks at proton colliders\footnote{
Single leptoquark production in various other high energy reactions was
studied in refs.~\cite{SINGLE}.}
proceeds either through the reaction
$q + g \rightarrow \Phi + e$~\cite{SCAL1} or
through charged lepton--quark (antiquark) fusion~\cite{ZER}.
For
the latter process the electron in the initial state
is provided by $e^+e^-$
pair production  of a photon which is collinearly radiated off a
quark of the second proton (antiproton).

The cross sections for single
scalar and vector-leptoquark production in $qg$-fusion
are~\cite{DNR} $\sigma \approx 0.7 ... 1.3~fb$
for a mass $M \sim 200 \GeV$ and
$(\lambda/e)\sqrt{b} = 0.075$.
For the second process, the production cross section for
leptoquarks at a mass $M \sim 200 \GeV$ is
given by
\begin{equation}
\sigma_{\Phi} \approx 0.074 pb \times
 \left ( \frac{\lambda}{e} \right )^2 J(\Phi)~b(\Phi)~.
\end{equation}
For values of $(\lambda/e)\sqrt{b} = 0.075$ the cross section amounts to
$\sigma = 0.4~fb$. These cross sections are too small to be measured
at TEVATRON currently.

Complementary to single leptoquark production~\footnote{One may also
search for virtual leptoquark exchange in
$e^+e^- \rightarrow q\overline{q}$ and
$q \overline{q} \rightarrow e^+e^-$ scattering at LEP2 and at TEVATRON,
respectively. Since $\lambda_{S,V} \ll 1$, the dominant contributions
come from the interference terms between the respective $s$-channel
contributions and leptoquark exchange in the $t$ or $u$ channel,
cf.~\cite{IND}. At the current luminosities it is not possible~\cite{DNR}
to constrain the fermionic couplings further by these reactions
in the mass range $M \sim 200 \GeV$.},
which relies
on the fermionic couplings of leptoquarks, one may consider the pair
production~\footnote{Pair production
cross sections for $e^+e^-$ annihilation, $ep-$, $\gamma \gamma-$ and
$e\gamma$ scattering
were calculated in
refs.~\cite{PAIR1}. For a survey on the search potential,
see      ref.~\cite{BB94}.} of scalar and vector
leptoquarks
at hadron colliders.
Even if the fermionic couplings of leptoquarks are small,
$\lambda/e \ll 1$, leptoquarks
may be produced through quark--antiquark and
gluon--gluon fusion  since the couplings to the gluon are well
determined. The pair production cross sections were calculated in
refs.~\cite{SCA,BBK}\footnote{
For a critical
discussion of other results,~refs.~\cite{SCAA},
see ref.~\cite{BBK}.}
for scalar and in refs.~\cite{VEC1,VEC2,BBK}
for vector leptoquarks by different
methods. While in refs.~\cite{VEC1} only the case of vector leptoquarks
carrying a Yang--Mills type coupling and in refs.~\cite{VEC2}
as well the case of
the minimal vector coupling
to the gluon were considered,  the  general case
of
anomalous couplings, $\kappa_G$ and $\lambda_G$, of vector leptoquarks
to the gluon was delt with in ref.~\cite{BBK}.

In figure~1  the total
pair production cross section is shown for scalar
leptoquarks using the CTEQ3 (LO) parametrization~\cite{CTEQ3} for
the parton densities.
For $M \sim 200~\GeV$ the scalar pair production cross section
amounts to $\sigma(M \sim 200 \GeV) = 0.1~pb$, taking $\mu =
\sqrt{\hat{s}}$
both as the factorization and renormalization scale.
Other
choices as $\mu = M_S$ lead to a  cross section which is larger by a
factor of $\sim 1.6$ for $M \sim 200 \GeV$, showing the scale dependence
in lowest order.\footnote{
A discussion on this aspect with M. Mangano is gratefully
acknowledged.}

The pair production cross section for
vector leptoquarks depends sizeably
on the anomalous couplings $\kappa_G$ and $\lambda_G$ of the vector
leptoquarks to the gluon, see figure~1.
If the coupling would be of the Yang--Mills type, i.e.
$\kappa_G = \lambda_G = 0$, the event rate could reach
about 1000 events for $M \sim 200 \GeV$ at each TEVATRON
experiment. Even in the case of the minimal vector coupling,
$\kappa_G = 1$ and $\lambda_G = 0$, the  rate amounts to 100 events.
Both these cases are likely to be
excluded due to the event rates measured
at the TEVATRON experiments. However, since a prediction on the
vector couplings $\kappa_G$ and $\lambda_G$ does not exist,  a
global limit can only be derived
minimizing the integrated cross section for these parameters.
For a mass of
$M \sim 200 \GeV$ one obtains
\begin{equation}
\sigma_V^{min}(M = 200 \GeV)
 = 0.20~pb~(Br(\Phi_V \rightarrow e q))^2
\end{equation}
for events with a $2e~2jet$ final state assuming
$\mu = \sqrt{\hat{s}}$.
For $\mu = M_V$ $\sigma^{min}(M = 200 \GeV) =
0.29~pb~(Br(\Phi_V \rightarrow e q))^2$  is obtained.
The corresponding
anomalous couplings are $\kappa_G = 1.3$ and $\lambda_G = -0.34$.

These leading order
estimates show that the pair production process of leptoquarks
may
yield up to 10 .. 16 signal events at an integrated luminosity of
${\cal L} = 100~pb^{-1}$
for scalar leptoquarks and, in the
presence of only the two anomalous couplings $\kappa_G$ and $\lambda_G$,
for vector leptoquarks at the minimal cross section,
                       up to 20 .. 30  events, depending on the
value of the factorization and renormalization scale~\footnote{A complete
calculation of the $O(\alpha_s)$ K-factor both for scalar and vector
leptoquark pair production is not yet available.}
and the yet unknown branching ratios.
The detection of the $2e~2jet$ final state is complicated due to
different background reactions, see e.g.~\cite{BGR}, which can only
be reliably studied by the experiments themselves.
Given the uncertainty in the branching fraction $Br(\Phi \rightarrow
eq)$, both scalar and
 vector leptoquarks seem
to be not yet  excluded by the TEVATRON
data.
On the basis of the present statistics bounds on the anomalous couplings
$\kappa_G$ and $\lambda_G$ for vector leptoquarks
can be derived
up to masses of $O(300~\GeV)$.
\section{Conclusions}
An interpretation of the recently reported
                                        excess of events in the range of
high $x$ and $Q^2$ at HERA in terms of single
leptoquark production yields
leptoquark-fermion couplings, $\lambda_{f_i, f_j}$, which are
compatible with other limits, particularly those obtained from low energy
reactions, for $\lambda_{e^+ u}$ and $\lambda_{e^+ d}$.
Still more statistics is needed to clearly establish an effect.

If the observed excess of events is due to leptoquarks, these states
should as well be detected in the reaction channel $e^+ q \rightarrow
g \Phi_{S,V}$ at HERA at
an increased
integrated luminosity.
The measurement of this reaction  allows to
investigate the coupling of the state $\Phi$ to gluons and provides an
independent possibility to determine its spin.

The FERMILAB experiments, at a
higher statistics, will be able to
either clearly confirm or exclude the
interpretation of the current excess of events found at HERA as being
due to
leptoquark production. For vector leptoquarks already at
the current statistics bounds on the range of the couplings $\kappa_G$
and $\lambda_G$ can be derived.
                                It is likely that both Yang-Mills type
and minimal couplings of the leptoquarks to the gluon are already exluded
by the observed event rates.
On the other hand the TEVATRON data do not exclude
scalar and vector leptoquarks, for some range
in $\kappa_G$ and $\lambda_G$, at a mass $M \sim 200 \GeV$.

The search for leptoquark pair production
at TEVATRON
is  complementary to the search for single leptoquark production at HERA
both
due to the
couplings involved in the production mechanism, as well as with respect
to the alternative
possibility
to investigate the spin of the produced state in case of a confirmation
of the effect.

\vspace{2mm}
\noindent
{\bf Acknowledgement.}~It is a pleasure for me to thank
C. Berger, E. Boos, W. Buchm\"uller, R.~Eichler,
M. Klein, A. Kryukov, E. Reya, R. R\"uckl,
G. Wolf and P.M. Zerwas for discussions on the present topic during
the last years. I would like to thank F. Jegerlehner, M. Klein,
P. S\"oding and A. Wagner for reading the manuscript.

\vspace{2mm}
\noindent
{\sf
Note added.}~After completion of this paper we received the
papers~\cite{PAP}, in which
similar analyses were carried out, which are found to be in mutual
agreement.
\noindent

\newpage
\begin{center}

\mbox{\epsfig{file=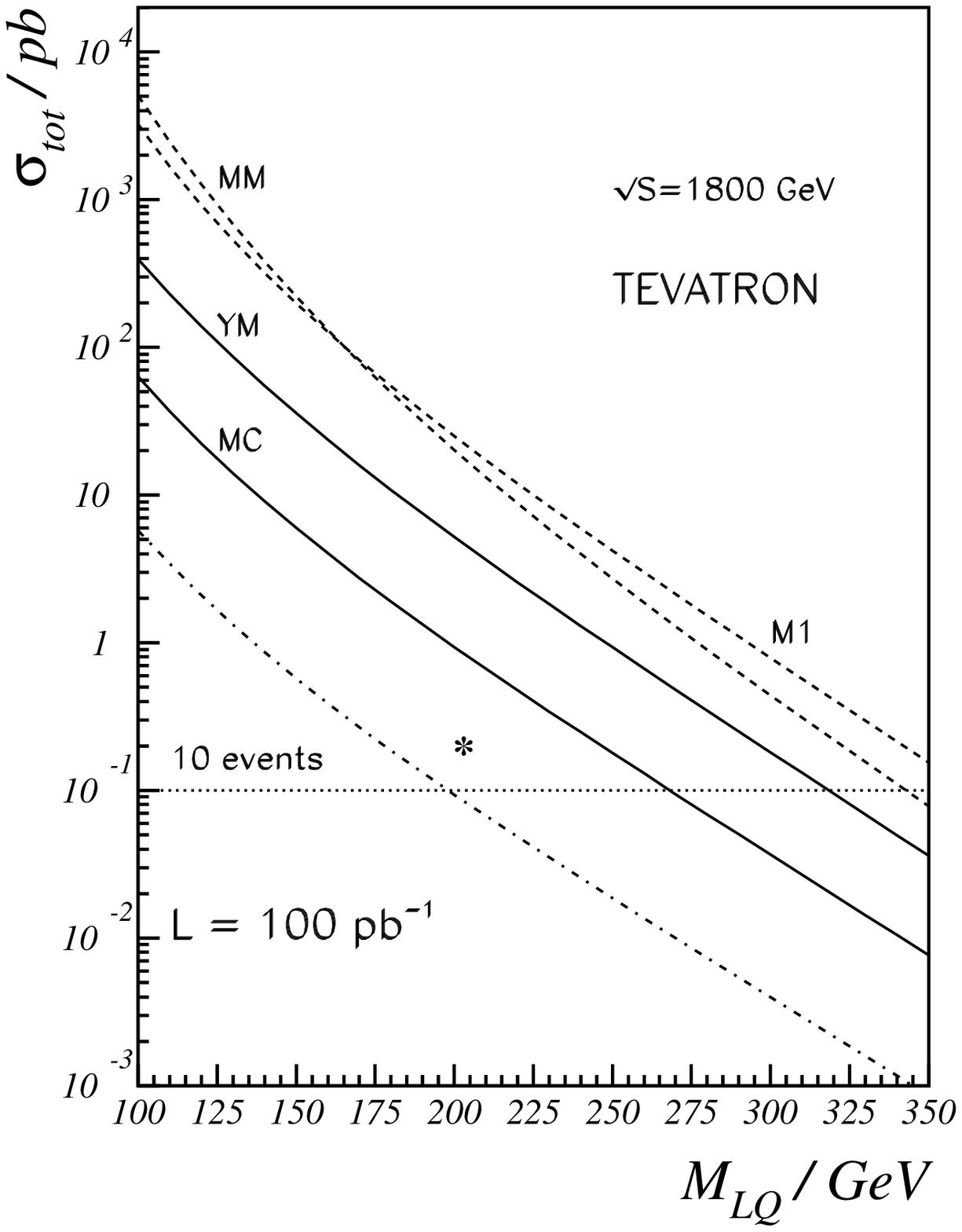,height=18cm,width=16cm}}

\vspace{2mm}
\noindent
\small
\end{center}
{\sf
Figure~1:~Integrated cross
sections for scalar
vector leptoquark pair production at the
TEVATRON,
$\sqrt{s}~=~1.8~\TeV$ choosing the renormalization and factorization
scale by $\mu = \sqrt{\hat{s}}$.
Dash-dotted line~: scalar leptoquarks; fill lines~:~YM~:
 Yang-Mills type coupling
$\kappa_G = \lambda_G  \equiv
0$;~MC~: minmal vector
coupling $\kappa_G = 1, \lambda_G = 0$; dashed lines~:~other choices
for the anomalous couplings~:~MM~:
$\kappa_G = \lambda_G = -1$,~M1~:
$\kappa_G = -1,\lambda_G = +1$. The asterisk denotes the minimum of
the pairproduction cross section for vector leptoquarks with
respect to the anomalous couplings at $M_V = 200~\GeV$.
}
\normalsize
\end{document}